\documentclass[12pt]{article}
\usepackage{a4wide,amsfonts,epsfig}
\input epsf
\newcommand{\tr}{\mbox{trace}\,}
\newcommand{\trs}{\mbox{trace}_2\,}

\newcommand{\ns}{\ \indent}
\newcommand{\news}{\setcounter{equation}{0}}

\newcommand{\sn}[2]{\,\mbox{sn}_#1\/#2}
\newcommand{\cn}[2]{\,\mbox{cn}_#1\/#2}
\newcommand{\dn}[2]{\,\mbox{dn}_#1\/#2}

\newcommand{\alphabf}{\mbox{\boldmath $\alpha$}}
\newcommand{\betabf}{\mbox{\boldmath $\beta$}}

\begingroup\makeatletter\ifx\SetFigFont\undefined
\def\x#1#2#3#4#5#6#7\relax{\def\x{#1#2#3#4#5#6}}%
\expandafter\x\fmtname xxxxxx\relax \def\y{splain}%
\ifx\x\y   
\gdef\SetFigFont#1#2#3{%
  \ifnum #1<17\tiny\else \ifnum #1<20\small\else
  \ifnum #1<24\normalsize\else \ifnum #1<29\large\else
  \ifnum #1<34\Large\else \ifnum #1<41\LARGE\else
     \huge\fi\fi\fi\fi\fi\fi
  \csname #3\endcsname}%
\else
\gdef\SetFigFont#1#2#3{\begingroup
  \count@#1\relax \ifnum 25<\count@\count@25\fi
  \def\x{\endgroup\@setsize\SetFigFont{#2pt}}%
  \expandafter\x
    \csname \romannumeral\the\count@ pt\expandafter\endcsname
    \csname @\romannumeral\the\count@ pt\endcsname
  \csname #3\endcsname}%
\fi
\fi\endgroup

\begin{document}

\title{\vskip -70pt
  \begin{flushright}
    {\normalsize{DAMTP-1999-104\\ SNUTP-99-042\\ KIAS-P99089\\ hep-th/9909218}}
  \end{flushright}
  \vskip 15pt
  {\bf \Large \bf Nahm Data and the Mass of 1/4-BPS States}
  \vskip 10pt}
\author{ 
  Conor J. Houghton,\thanks{E-mail: houghton@cuphyb.phys.columbia.edu}
\\[5pt] 
{\normalsize {\sl DAMTP, University of Cambridge, Silver St.,
      Cambridge, CB3 9EW, UK.}}
\\{\normalsize {\sl and}} \\
{\normalsize {\sl Physics Department, Columbia University, New York, New York 10027, USA.}\thanks{Address during academic year 1999/2000}}
\\[12pt]
Kimyeong Lee,\thanks{E-mail: kimyeong@phya.snu.ac.kr } \\[5pt]
  {\normalsize {\sl Physics Department and Center for Theoretical Physics,}}\\ 
  {\normalsize {\sl Seoul National University, Seoul 151-742, Korea.}}
\\
{\normalsize {\sl and}} \\
{\normalsize {\sl School of Physic, Korea Institute for Advanced
  Study}} \\
{\normalsize {\sl 207-43, Cheongryangryi-Dong, Dongdaemun-Gu, Seoul
  130-012, Korea.\thanks{Address from September 1, 1999} }} \\
}

\date{September 29, 1999}
\maketitle
\begin{abstract}
  
The mass of 1/4-BPS dyonic configurations in ${\cal N}=4$ $D=4$
supersymmetric Yang-Mills theories is calculated within the Nahm
formulation.  The $SU(3)$ example, with two massive monopoles and one
massless monopole, is considered in detail. In this case, the
massless monopole is attracted to the massive monopoles by a linear
potential.

\end{abstract}

\newpage

\section{Introduction\label{intro}}
\news

\ns In the context of ${\cal N}=4$ supersymmetric Yang-Mills theories in
four-dimensional spacetime, BPS magnetic monopoles are referred to as
1/2-BPS states, because they are invariant under half of the
supersymmetry. Recently, 1/4-BPS states have also been
considered. 1/4-BPS states are invariant under only a quarter of the
supersymmetry and form somewhat larger supermultiplets.

Generally, there are six scalar fields in ${\cal N}=4$ supersymmetric
Yang-Mills theory. For a configuration to be 1/4 BPS, all but two of
these scalar fields must vanish and the remaining two must satisfy two
field equations called BPS equations. The first of these equations
requires that the gauge fields $A_i$ and one of the scalar fields,
$b$, must satisfy the usual Bogomolny equation for BPS monopoles. The
second BPS equation requires that the other scalar field, $a$, must
satisfy the covariant Laplace equation in the background of the
solution, $A_i$ and $b$, of the first BPS equation.  

A point in the moduli space of 1/2-BPS configurations corresponds to a
unique 1/4-BPS configuration; the field $a$ is determined uniquely.
This means the contribution of $a$ to the mass of the 1/4-BPS state is
a potential function over the moduli space. The contribution of $a$ to
the mass is referred to as the electric part of the mass, or simply, the
electric mass. It is thought that there may be a moduli space
approximation to the low energy dynamics of 1/4-BPS states with
kinetic term given by the usual moduli space metric and with potential
term given by half  the electric mass.

It is difficult to solve the two BPS equations. The most tractable
approach is to employ the Nahm formulation. Using the Nahm formulation,
the fields were found for a simple case in \cite{LY}. Another approach
is to study spherically symmetric solutions and use a spherical ansatz
to solve the field equations \cite{Betal1,IS}. 

However, past experience has shown that a great deal can be learned
about 1/2-BPS configurations, without knowing the explicit fields. It
appears that this is also the case with 1/4-BPS configurations.  The
solutions of the first BPS equations are described by moduli space
coordinates and there is a natural metric on the moduli space. In a
number of examples, this metric is known even though the fields are
not. Furthermore, the electric charge and the electric part of the BPS
energy can be obtained from the moduli space metric without solving
the BPS equations for the fields \cite{T}.

In this paper, we show that the electric mass may also be calculated
directly from the Nahm data, without having to calculate either the
fields or the metric.  We apply this to two $SU(3)$ examples. In the
first example, $b$ is a $(1,1)$-monopole. In this case, the electric
mass is already know from more direct calculations \cite{LY,T}.  In
the second example, the $b$ field is the $(2,[1])$-monopole considered
by Dancer \cite{D1,D2}. The electric mass has not been previously
calculated for this case and, so, it is discussed in some detail.

A $(2,[1])$-monopole is a monopole in the theory with $SU(3)$ symmetry
broken to $U(2)$ by the asymptotic value of $b$. In other words,
$SU(3)$ has two simple roots, $\alphabf$ and $\betabf$, and the
asymptotic value of $b$ is perpendicular to $\betabf$, permitting a
$U(2)$ of gauge inequivalent large gauge transformations of the
fields. This means that the theory has two types of monopoles: massive
$\alphabf$ monopoles and massless $\betabf$ monopoles. The nonabelian
case is usefully thought of as a limit of the generic case, where the
asymptotic value of $b$ is not perpendicular to either root
\cite{LWY3}.

A $(2,[1])$-monopole is composed of two $\alphabf$ monopoles and one
$\betabf$ monopole.  The moduli space of $(2,[1])$-monopoles describes
the solutions of the first BPS equation. The net magnetic charge is
purely abelian and the massless $\betabf$ monopole forms a nonabelian
cloud surrounding the two massive $\alphabf$ monopoles.  However, this
discussion refers only to the contribution that $b$ makes to the mass.
The asymptotic value of the second Higgs field, $a$, breaks the
symmetry further to $U(1)\times U(1)$. This symmetry breaking pattern
is required for genuine 1/4-BPS configurations. When the asymptotic
values of the two Higgs fields are proportional, the Higgs fields are
proportional everywhere. This means that there is really only one
active scalar field and the configuration is 1/2 BPS.  In a genuine
1/4-BPS configuration, the $\betabf$ monopole is not massless if the
contribution of the $a$ field is included.

A similar situation arose in the $(1,[1],1)$ case recently considered
by one of us $(KL)$ \cite{lee}. Here, the $SU(4)$ symmetry is broken
to $U(1) \times SU(2) \times U(1)$ by the $b$ field and is broken
maximally by the $a$ field.  Due to the recent progress in the
understanding of 1/4-BPS configurations, some remarks can now be made
about the $(2,[1])$ case which also apply to the $(1,[1],1)$ case. It
is emphasized that the precise position of the massless $\betabf$
monopole is important in the 1/4-BPS configuration. As far as the field $b$
is concerned, the position of the massless monopole can be transformed
by the unbroken symmetry. In fact, if a massless monopole is
considered as the massless limit; then some of the moduli of the monopole
become, in the limit, parameters of the orbit of the unbroken
symmetry.  On the other hand, the asymptotic $a$ field breaks the
$SU(2)$ symmetry and so the solution depends on the position of the
massless monopole. Similarly, if there were several identical
massless monopoles, the solution of the second BPS equation would
depend on the moduli of the massless monopoles.

Recently, a low energy effective lagrangian for the moduli dynamics of
1/4-BPS configurations has been found \cite{Betal2}. It consists of a
kinetic part and a potential part. The kinetic part is given by the
moduli space metric of the corresponding 1/2-BPS monopole. The
potential part is given by half of the electric mass. The
non-relativistic lagrangian has a BPS bound, and the BPS configuration
saturating this bound turns out to be the 1/4-BPS field theoretic
configuration.  Naive criteria, for specifying the valid region of
this low energy lagrangian, require that the kinetic energy and the
potential energy are much smaller than the rest mass of the monopoles
involved. When massless monopoles are involved, it is not clear
whether there is any valid region. However, if such a region exists,
it would have to be one where the energy is smaller than the magnetic
mass of the configuration. In this paper, we assume that a valid
region exists and we explore the consequences. The picture which
emerges appears to be consistent.

The potential in the $(2,[1])$ case is quite interesting. It is
repulsive and the relative electric charge between dyons generates an
effective attractive force. There is a minimum allowed electric
charge, which matches nicely with the string picture. The balance of
the attractive electric force and the repulsive force leads to the BPS
configurations. Moreover, when we try to remove the apparently
massless $\betabf$ monopole away from the $\alphabf$ monopoles, it
turns out that the potential grows linearly with distance from the two
$\alphabf$ monopoles to the $\betabf$ monopole. This is a sort of
confinement.  Of course, it would be strange if we could take out the
massless monopole with a finite energy cost since it would then appear
to be both massive and massless.

In section \ref{revBPS}, there is a review of the physics of 1/4-BPS
configurations. In section \ref{Nfsec}, we discuss 1/4-BPS
configurations in the Nahm formulation and derive a formula for the
electric mass. This formula is used to calculate the known electric
mass potential in the $(1,1)$ case.  In section \ref{dsect}, the
formula is used to calculate the mass of 1/4-BPS states in the
$(2,[1])$ case. A string configuration equivalent to this state is
proposed.  Finally, we conclude with some remarks in section
\ref{dissec}.

\section{1/4-BPS configurations \label{revBPS}}
\news

\ns A ${\cal N}=4$ supersymmetric Yang-Mills theory has a six
component scalar field. Four of these components are zero in 1/4-BPS
configurations \cite{LY} and it is possible to choose two independent
orthogonal Higgs fields $a$ and $b$ satisfying
\begin{equation}
B_i=D_i b
\label{firstbps}
\end{equation}
and
\begin{equation}
D_iD_i a+[b,[b,a]]=0\label{CL}
\end{equation}
where the coupling constant, $e$, is set to one. In this context, these
are referred to as the first and second BPS equations.  The gauge group
is $SU(N)$. Thus, $A_i$ and $b$ satisfy the usual Bogomolny equation
for 1/2-BPS monopoles and $a$ satisfies a covariant Laplace equation
in the background of $A_i$ and $b$. The equation satisfied by $a$ is
the same as the equation satisfied by a large gauge transformation of
$A_i$ and $b$ obeying the background gauge condition. The appearance
of the Bogomolny equation apparently reflects the fact that some of
the supersymmetry is unbroken. There is no Bogomolny equation in the
non-BPS case, where three of the scalar fields are active \cite{IS}.
  
A general configuration has both magnetic and electric
charges. Asymptotically, the Higgs field lies in the gauge orbit of
\begin{eqnarray}
&&b\simeq {\bf b}\cdot {\bf H}-\frac{1}{4\pi r} {\bf g}\cdot {\bf H},\\
&&a\simeq {\bf a}\cdot {\bf H}-\frac{1}{4\pi r} {\bf q}\cdot {\bf H},
\label{asympa}
\end{eqnarray}
where the dot product of bold quantities is in the Cartan space. The
mass of the corresponding configuration is
\begin{equation}
{\bf b}\cdot {\bf g} + {\bf a}\cdot {\bf q}
\end{equation}
Thus, there are two contributions to the mass: ${\bf b}\cdot {\bf g}$
and ${\bf a}\cdot {\bf q}$. These are referred to as the magnetic mass
and the electric mass respectively.

The solutions of the first BPS equation, (\ref{firstbps}), are 1/2-BPS
monopoles. Generally, the asymptotic value, ${\bf b}$, of $b$ breaks
$SU(N)$ to $U(1)^{N-1}$ and there exists a unique set of simple roots
$\betabf_1, \betabf_2, \ldots, \betabf_{N-1}$ such that
$\betabf_I\cdot {\bf b}>0$ for $I=1\ldots N-1$. For each simple root,
there is a fundamental monopole with four zero modes. For any magnetic charge
\begin{equation}
{\bf g} = g (k_1 \betabf_1 + k_2
\betabf_2+...+k_{N-1}\betabf_{N-1})
\end{equation}
with non-negative $k_I$, there exists a family of 1/2-BPS solutions of
the first BPS equation, called
$(k_1,k_2,\ldots,k_{N-1})$-monopoles~\cite{weinberg}. Usually
$g=4\pi$. The solutions 
are uniquely characterized by their moduli space coordinates. Thus,
any specific solution of the first BPS equation is given by the
coordinates ${z^p}$ of the moduli space of these 1/2-BPS
monopoles. The dimension of the moduli space is $4 \times
(k_1+k_2+...+k_{N-1})$ and there exists a naturally defined metric on
this moduli space, given by the $L^2$ norm of gauge-orthogonal field
variations.  The magnetic mass is the same for all monopoles with the
same magnetic charge and is given by
\begin{equation}
{\bf b}\cdot {\bf g}=g\sum_I k_I \mu_I
\end{equation}
where $g\mu_I$ is the mass of the $I$th type of monopole.

The symmetry breaking is not maximal when $\betabf_I\cdot {\bf b}=0$
for some $I$. The corresponding fundamental monopole becomes massless
and does not exist in isolation. However, when the total magnetic
charge ${\bf g}$ is purely abelian, so that ${\bf g}\cdot \betabf_I
=0$ for $\betabf_I\cdot {\bf b}=0$, there are massless monopoles with
$k_I = (k_{I+1}+k_{I-1})/2 $.  These massless monopoles appear as a
nonabelian cloud surrounding the massive monopoles. As long as the
total magnetic charge remains purely abelian the dimension of the
moduli space does not change in the massless limit
\cite{weinberg,LWY3}. However, the meaning of the moduli space coordinate
changes from the point of view of 1/2-BPS configurations. The position
and phase of massless monopoles become the unbroken nonabelian gauge
orbit parameters and gauge invariant cloud parameters.

A solution of the second BPS equation can be found for each
solution of the first BPS equation. In fact, the second BPS
equation is identical to the zero mode equation for gauge-orthogonal
large gauge transformations of the fields. There are $N-1$ such zero
modes since $SU(N)$ breaks into $N-1$ abelian $U(1)$ groups. From the
solution of the second BPS equations, the electric charges carried by
the $k_I$ $\betabf_I$ monopoles can be read off from the asymptotic
field (\ref{asympa}).

Thus, when the asymptotic value, ${\bf b}$, leaves some of the
nonabelian gauge symmetry unbroken, the 1/2-BPS configurations may
involve massless monopoles. The asymptotic value ${\bf a}$ may
preserve the unbroken symmetry by ${\bf b}$ or break it further. When
there is an unbroken nonabelian symmetry group, there may be
non-vanishing, nonabelian electric charge for a 1/4-BPS
configurations. This is shown in a simple case in \cite{lee}.

\section{The Nahm formulation \label{Nfsec}}
\news

\ns In this section, we first review the Nahm formulation for the first
BPS equation. We then derive a formula for the electric mass using the
Nahm formulation. There is a complication when two adjacent charges
are equal, as in the $(1,1)$ case. In subsection \ref{eacsec}, we extend
our formula to these cases. The specific example of the $(1,1)$ case
is considered in subsection \ref{11case}. In the next section, the
formula is applied to the $(2,[1])$ case.

BPS monopoles are classified by Nahm data \cite{N,Hi}.  Nahm data are
a 4-vector of skewhermetian matrix functions of one variable over the
subdivided interval defined by the eigenvalues of the asymptotic Higgs
field ${\bf b}\cdot{\bf H}$. Inside each subinterval, the data satisfy
the Nahm equation
\begin{equation}
\frac{dT_i}{ds}+[T_0,T_i]=[T_j,T_k]
\end{equation}
where $(i\,j\,k)$ is a cyclic permutation of $(1\,2\,3)$. Each
subinterval corresponds to one of the unbroken $U(1)$ subgroups of
$SU(N)$ and the magnetic charge around that $U(1)$ determines the size
of the Nahm matrices over that subinterval. Thus, a
$(k_1,k_2,\ldots,k_{N-1})$-monopole with Higgs field at spatial
infinity is given by
\begin{equation}
{\bf b}\cdot{\bf H}=-i\mbox{diag}(s_1,s_2,\ldots,s_N)
\end{equation}
where $s_1<s_2<\ldots<s_N$, corresponds to Nahm data over the
interval $(s_1,s_N)$. Over $(s_1,s_2)$, the Nahm matrices are
$k_1\times k_1$, over $(s_2,s_3)$ they are $k_2\times k_2$ and so on.
It is useful to illustrate Nahm data with a graph,
taking the value $k_I$ over the $I$th interval.

Boundary conditions relate the Nahm matrices in abutting subintervals.
For
\begin{equation}
\begin{array}{c}
\begin{picture}(205,85)(55,605)
\thinlines
\put( 60,620){\line( 1, 0){200}}
\put(100,680){\line( 1, 0){ 60}}
\put(160,680){\line( 0, -1){ 20}}
\put(160,660){\line( 1, 0){ 70}}
\multiput(160,660)(0,-8){6}{\line(0,-1){  4}}
\multiput(100,680)(-8,0){3}{\line(-1,0){  4}}
\multiput(230,660)( 8,0){3}{\line( 1,0){  4}}
\put( 95,623){\vector(0,1){  56}}  
\put( 95,623){\vector(0,-1){  0}} 
\put(160,620){\line( 0,-1){  5}}
\put( 81,648){\makebox(0,0)[lb]{\smash{\SetFigFont{12}{14.4}{rm}$k_1$}}}
\put(157,605){\makebox(0,0)[lb]{\smash{\SetFigFont{12}{14.4}{rm}$s_0$}}}
\put(240,638){\makebox(0,0)[lb]{\smash{\SetFigFont{12}{14.4}{rm}$k_2$}}}
\put(235,623){\vector(0,1){  36}}
\put(235,623){\vector(0,-1){  0}}
\end{picture}\end{array}
\end{equation}
the Nahm matrices are $k_1\times k_1$ matrices if
$s<s_0$ and $k_2\times k_2$ matrices if $s>s_0$. The boundary
condition requires that
\begin{equation}
T_i(s_0-)=\mbox{blockdiag}\left(R_i/(s-s_0),T_i(s_0+)\right)\label{frombelow}
\end{equation}
where the $(k_1-k_2)\times(k_1-k_2)$ residue matrices $R_i$  must form
the $(k_1-k_2)$-dimensional irreducible 
representation of $\mathfrak{su}(2)$. Thus, part of the data carries
through the junction and the rest has a pole with residues of a
particular type.

The boundary conditions are different when $k_1=k_2$. In this case the data
carrying though may have a rank one discontinuity. Thus, for
\begin{equation}
\begin{array}{c}
\begin{picture}(205,65)(55,605)
\thinlines
\put( 60,620){\line( 1, 0){200}}
\put(100,660){\line( 1, 0){ 60}}
\put(160,660){\line( 1, 0){ 70}}
\multiput(160,660)(0,-8){6}{\line(0,-1){  4}}
\multiput(100,660)(-8,0){3}{\line(-1,0){  4}}
\multiput(230,660)( 8,0){3}{\line( 1,0){  4}}
\put( 95,623){\vector(0,1){  36}}  
\put( 95,623){\vector(0,-1){  0}} 
\put(160,620){\line( 0,-1){  5}}
\put( 84,638){\makebox(0,0)[lb]{\smash{\SetFigFont{12}{14.4}{rm}$k$}}}
\put(157,605){\makebox(0,0)[lb]{\smash{\SetFigFont{12}{14.4}{rm}$s_0$}}}
\end{picture}\end{array}
\end{equation}
the boundary condition at $s=s_0$ is
\begin{equation}
\Delta T_i=T_i(s_0+)-T_i(s_0-)=\frac{i}{2}\trs{\sigma_i w w^\dagger}
\end{equation}
where $w$ is a $2\times k$ matrix, $ww^\dagger$ is a tensor product
over both indices and $\trs(\cdot)$ sums the diagonal of the
$2\times2$ part of the product. $w$ is often called the jump or
matching data. The boundary condition is sometimes included in the
Nahm equation as a source term
\begin{equation}
\frac{dT_i}{ds}+[T_0,T_i]=[T_j,T_k]+\frac{i}{2}\trs{\sigma_i w w^\dagger}\delta(s-s_0).
\end{equation}
This practice is not adopted here.

The boundary conditions for non-maximal symmetry breaking can be
derived from those above by identifying eigenvalues.

There is a group action on Nahm data given by 
\begin{eqnarray}\label{Ndgact}
T_0&\rightarrow&G T_0 G^{-1}-\frac{dG}{ds}G^{-1},\nonumber\\
T_i&\rightarrow&G T_i G^{-1}
\end{eqnarray} 
where $G$ is a group valued function of $s$.  For this to act on Nahm
data, $G$ must satisfy certain boundary conditions on the subintervals.
Depending on how strong the boundary conditions satisfied are, $G$ is a
large or small gauge transformation. The moduli space of Nahm data is
defined relative to small gauge transformations. The precise form of
boundary conditions on the group transformations and the criterion
distinguishing large and small gauge transformations of the Nahm data
are discussed for specific examples in \cite{D1,HIM}.

In \cite{LY}, the Nahm formulation of 1/4-BPS states is derived by
applying the Fourier analysis methods of \cite{KvB} to the ADHM
formulation of the Laplace equation for an instanton background
\cite{O}. By partially performing the ADHMN construction ${\bf a}\cdot
{\bf q}$ is calculated for $(1,1)$-monopoles. A technique is now presented
which simplifies the calculation of the mass in the Nahm
formulation. This technique relies on the isometry between the space of
Nahm data and the moduli space of monopoles. In fact, while it is
believed that these spaces are isomorphic in general, this has only
been proven in specific cases \cite{Nak}.

In \cite{T}, an expression is derived for the mass of a 1/4-BPS state
as a function over the monopole moduli space. The argument that is used there
is reversed to allow us to write the mass in terms of the solution to
the covariant Laplace equation on Nahm data. The electric contribution
to the mass is
\begin{eqnarray}\label{lq}
{\bf a}\cdot{\bf q}&=&\tr \int d^3x \sum_i\partial_i(a D_ia)\nonumber\\
                   &=&\tr \int d^3x \{\sum_i(D_ia)^2-[a,b]^2\}.
\end{eqnarray}
In \cite{T}, it is noted that $D_ia$ is a large gauge transformation in $A_i$
and can be written as 
\begin{equation}
D_ia=\sum_p{\bf a}\cdot {\bf K}^p\delta_pA_i.
\end{equation}
In the same way,
\begin{equation}
-i[b,a]=\sum_p{\bf a}\cdot {\bf K}^p\delta_pb
\end{equation}
where $\delta_p A_i$ and $\delta_p b$ are the zero modes for the 1/2-BPS configurations and ${\bf K}^p$ are the Killing vectors for the large gauge transformations.
Substituting these expressions into (\ref{lq}) and using the equation
satisfied by $a$ gives the Tong formula
\begin{equation}\label{Tf}
{\bf a}\cdot{\bf q}=\sum_{p,q}g_{pq}({\bf a}\cdot {\bf K}^p)({\bf a}\cdot {\bf K}^q)
\end{equation}
where 
\begin{equation}
g_{pq}(z) =\tr\int d^3x\left(\sum_i \delta_pA_i\delta_qA_i+\delta_p b\,\delta_q b\right).
\end{equation}
This formula allows the electric mass to be calculated from the
metric; it does not require that the fields are known.

The metric can also be written in terms of Nahm data:
\begin{equation}\label{metN}g_{pq}=-g\tr\int d s\sum_\mu \delta_pT_\mu\delta_qT_\mu.
\end{equation}
This expression can be substituted into the Tong formula (\ref{Tf})
and the derivation can be reversed to give a formula for ${\bf
a}\cdot{\bf q}$ in terms of a large gauge transformation of the Nahm
data satisfying the background gauge condition. There is a factor of
$g$ in (\ref{metN} because the mass of a static monopole is $g\mu$
rather than $\mu$.

An infinitesimal gauge transformation $h$ of the Nahm data
is given by 
\begin{eqnarray}\label{gatra}
\delta T_0&=&-\frac{dh}{ds}-[T_0,h],\nonumber\\
\delta T_i&=&-[T_i,h]
\end{eqnarray}
and the background gauge condition is
\begin{equation}
\frac{d\delta T_0}{ds}+\sum_\mu[T_\mu,\delta T_\mu]=0.
\end{equation}
This is derived by requiring that variations are orthogonal to small gauge
transformations of the data.
Thus, a gauge-orthogonal large gauge transformation, $p$, satisfies the covariant Laplace equation
\begin{equation}\label{NdcL}
[\frac{d}{ds}+T_0,[\frac{d}{ds}+T_0,p]]+\sum_i[T_i,[T_i,p]=0.
\end{equation}
This is the equation discussed in \cite{LY}: the only differences
result from the convention of using skewhermetian rather than
hermetian Nahm data. 

Now, substituting
\begin{eqnarray}
\sum_p {\bf a}\cdot {\bf K}^p \delta_p T_0 &=&-\frac{dp}{ds}-[T_0,p],\nonumber\\
\sum_p {\bf a}\cdot {\bf K}^p \delta_p T_i &=&-[T_i,p]
\end{eqnarray}
into (\ref{metN}) and using the covariant Laplace equation (\ref{NdcL}) gives
\begin{equation}\label{masform}
{\bf a}\cdot{\bf q}=-g\tr\int ds \frac{d}{ds}\left[p\left(\frac{dp}{ds}+[T_0,p]\right)\right]=-g\tr\int ds \frac{d}{ds}\left(p\frac{dp}{ds}\right).
\end{equation}
This is the formula for the mass in terms of the 1/4-BPS Nahm data. It
allows the ADHMN construction to be avoided when calculating the
mass. 

A difficulty with using this formula is calculating what ${\bf a}$
is. In the example considered in section \ref{dsect}, this is not
difficult as the form of the group action on the Nahm is very
clear. It is more difficult in more trivial examples, where $p$ is
proportional to the identity. In these cases it seems the ADHMN
construction must be examined. The ADHMN construction for 1/4-BPS
states is described in \cite{LY}. In short, a linear equation, known
as the ADHMN equation, is solved for a set of $N$ complex vector
functions $v_I(s;x_1,x_2,x_3)$. $b$ and $a$ are then given by
\begin{eqnarray}\label{ADHMconst}
b_{IJ}&=&\int is v_I^\dagger v_J ds\nonumber\\
a_{IJ}&=&\int  v_I^\dagger (p\otimes{\bf 1}_2) v_J ds.
\end{eqnarray}
There is a tensor product with ${\bf 1}_2$ in the formula for $a$.
This matches the tensor product in the ADHMN equation.

Let us consider the trivial $SU(2)$ example.  In the $SU(2)$
case, $b$ is a $k$-monopole and there is no genuine 1/4-BPS state,
since the only solution to the second BPS equation has $a$
proportional to $b$. The only non-singular solution to the covariant Laplace
equation (\ref{NdcL}) is
\begin{equation}
p= Ais{\bf 1}_k
\end{equation}
where $A$ is a constant. By (\ref{ADHMconst}) this means
\begin{equation}
a=A b.
\end{equation}
Thus ${\bf a}=A{\bf b}$, ${\bf q}=A{\bf a}$
and the electric mass can be calculated both directly and by using the
formula (\ref{masform}). Either way
\begin{equation}
{\bf a}\cdot{\bf q}=A^2gk\mu=A^2 {\bf b}\cdot{\bf g}. 
\end{equation}

\subsection{Equal adjacent charges: including the jump data\label{eacsec}}

\ns Recall that in the case where there are equal adjacent charges the
Nahm data is augmented by jump data. The jump data appear in the
metric: the metric on the
\begin{equation}
\begin{array}{c}
\begin{picture}(205,65)(55,605)
\thinlines
\put( 60,620){\line( 1, 0){200}}
\put(100,620){\line( 0, 1){ 40}}
\put(100,660){\line( 1, 0){ 60}}
\put(160,660){\line( 1, 0){ 70}}
\put(230,660){\line( 0,-1){ 40}}
\multiput(160,660)(0.00000,-8.00000){6}{\line( 0,-1){  4.000}}
\put( 95,623){\vector(0,1){  36}}  
\put( 95,623){\vector(0,-1){  0}} 
\put(230,620){\line( 0,-1){  5}}
\put(100,620){\line( 0,-1){  5}}
\put(160,620){\line( 0,-1){  5}}
\put( 84,638){\makebox(0,0)[lb]{\smash{\SetFigFont{12}{14.4}{rm}$k$}}}
\put( 97,605){\makebox(0,0)[lb]{\smash{\SetFigFont{12}{14.4}{rm}$s_1$}}}
\put(157,605){\makebox(0,0)[lb]{\smash{\SetFigFont{12}{14.4}{rm}$s_2$}}}
\put(227,605){\makebox(0,0)[lb]{\smash{\SetFigFont{12}{14.4}{rm}$s_3$}}}
\end{picture}\end{array}
\end{equation}
data is
\begin{equation}
g_{pq}=-g\tr\int_{s_1}^{s_3} ds\sum_\mu(\delta_pT_\mu\delta_qT_\mu)+g\tr (\trs \delta_{\{p}w \delta_{q\}}w^\dagger).
\end{equation}
The metric is used to calculate the background gauge condition. Under
small gauge transformations 
\begin{equation}
\delta w = h(s_2)w.
\end{equation}
Imposing gauge orthogonality gives the boundary condition 
\begin{equation}\label{bgbc}
\Delta\delta T_0 = \frac{1}{2}\left(\trs w\delta
  w^\dagger - \trs \delta w w^\dagger\right).
\end{equation}
Large gauge  transformations of $w$ allow
\begin{equation}
\delta w = p(s_2) w - q w
\end{equation}
where $q$ is a pure imaginary number. Substituting this into the
background gauge boundary condition (\ref{bgbc}) gives 
\begin{equation}\label{pjump}
\Delta\left(\frac{dp}{ds}+[T_0,p]\right)=\frac{1}{2}\{p,\trs ww^\dagger\}-q\,\trs ww^\dagger.
\end{equation}
Repeating the previous calculation with the jump data included 
\begin{equation}
{\bf a}\cdot{\bf q}=-g\tr\int d s
\frac{d}{ds}\left(p\frac{dp}{ds}\right)-g\tr[\trs(ww^\dagger)(p(s_2)-q)^2].
\end{equation}
Substituting the boundary condition(\ref{pjump}) into this formula gives 
\begin{equation}\label{emeac}
{\bf a}\cdot{\bf
  q}=g\left[p(s_1)\left.\frac{dp}{ds}\right|_{s_1}-p(s_3)\left.\frac{dp}{ds}\right|_{s_3}+q\Delta\left(\frac{dp}{ds}\right)\right].
\end{equation}
Thus, $q$ plays the same role at junctions with jump data as $p$ does
at end points.

As with the 1/2-BPS dyon above, to identify ${\bf a}$
the ADHMN construction must be examined. As described by Nahm \cite{N},
when there are equal adjacent
charges the complex vector functions $v_I(s;x_1,x_2,x_3)$ are
supplemented with jumping data $\rho_I$ where
\begin{equation}
\Delta v_I = \rho_I w
\end{equation}
and
\begin{equation}\label{ADHMNb}
b_{IJ}=is_2 \rho_I^\star \rho_J +\int_{s_1}^{s_3} is v_I^\dagger v_J ds.
\end{equation}
Standard arguments involving approximate solutions of the ADHMN equation are then used to prove \cite{N,Hi} that
\begin{equation}
{\bf b}\cdot{\bf H}=-i\left(\begin{array}{ccc}s_1&&\\&s_2&\\&&s_3\end{array}\right)
\end{equation}
In the 1/4-BPS case \cite{LY},
\begin{equation}\label{ADHMNa}
a_{IJ}=q \rho_I^\star \rho_J +\int_{s_1}^{s_3}  v_I^\dagger (p\otimes {\bf 1}_2) v_J ds.
\end{equation}
If $p\propto {\bf 1}_k$, then the same standard arguments prove that
\begin{equation}\label{aHeac}
{\bf a}\cdot{\bf
H}=-\left(\begin{array}{ccc}p(s_1)&&\\&q&\\&&p(s_3)\end{array}\right).
\end{equation}

\subsection{The $(1,1)$ $SU(3)$ dyon case \label{11case}}

\ns ${\bf a}\cdot{\bf q}$ for the $(1,1)$ $SU(3)$ dyon has been calculated
twice already \cite{LY,T} and it is illustrative to calculate it
again. The Nahm data are
\begin{equation}
\begin{array}{c}
\begin{picture}(205,45)(60,605)
\thinlines
\put( 60,620){\line( 1, 0){200}}
\put(100,620){\line( 0, 1){ 20}}
\put(100,640){\line( 1, 0){ 60}}
\put(160,640){\line( 0,-1){ 0}}
\put(160,640){\line( 1, 0){ 70}}
\put(230,640){\line( 0,-1){ 20}}
\multiput(160,640)(0.00000,-8.00000){3}{\line( 0,-1){  4.000}}
\put( 95,623){\vector(0,1){  15}}  
\put( 95,623){\vector(0,-1){  0}} 
\put(230,620){\line( 0,-1){  5}}
\put(100,620){\line( 0,-1){  5}}
\put(160,620){\line( 0,-1){  5}}
\put( 84,626){\makebox(0,0)[lb]{\smash{\SetFigFont{12}{14.4}{rm}$1$}}}
\put( 97,605){\makebox(0,0)[lb]{\smash{\SetFigFont{12}{14.4}{rm}$s_1$}}}
\put(157,605){\makebox(0,0)[lb]{\smash{\SetFigFont{12}{14.4}{rm}$s_2$}}}
\put(227,605){\makebox(0,0)[lb]{\smash{\SetFigFont{12}{14.4}{rm}$s_3$}}}
\put(192,607){\makebox(0,0)[lb]{\smash{\SetFigFont{12}{14.4}{rm}$\mu_2$}}}
\put(127,607){\makebox(0,0)[lb]{\smash{\SetFigFont{12}{14.4}{rm}$\mu_1$}}}
\put(162,616){\vector(1,0){  66}}
\put(162,616){\vector(-1,0){  0}}
\put(102,616){\vector(1,0){  56}}
\put(102,616){\vector(-1,0){  0}}
\end{picture}\end{array}
\end{equation}
Up to a choice of origin and of spatial and group orientation, the Nahm
data are
\begin{equation}
(T_0,T_1,T_2,T_3)=\left\{\begin{array}{ll}(0,0,0,iR)&s\in(s_1,s_2)\\(0,0,0,0)&s\in(s_2,s_3)\end{array}\right.
\end{equation}
with
\begin{equation}
w=(0,\sqrt{2R})
\end{equation}
where $R$ is a real number. The covariant Laplace equation is
\begin{equation}
\frac{d^2p}{ds^2}=0
\end{equation}
with the boundary condition
\begin{equation}
\Delta \left(\frac{dp}{ds}\right)=2R(p(s_2)-a_2).
\end{equation}
The solution is
\begin{equation}
p=\left\{\begin{array}{ll}ip_1(s-s_1)+ia_1 & s\in(s_1,s_2)\\ ip_2(s-s_3)+ia_3
    & s\in(s_2,s_3)\end{array}\right.
\end{equation}
where the boundary conditions imply
\begin{eqnarray}
p_1\mu_1+a_1&=&-p_2\mu_2+a_3\nonumber\\
            &=&a_2+\frac{1}{2R}(p_2-p_1)
\end{eqnarray}
and $q=ia_3$. Solving for $p_1$ and $p_2$ gives
\begin{eqnarray}
p_1&=&\frac{a_3-a_1+2(a_2-a_1)\mu_2R}{\mu_1+\mu_2+2\mu_1\mu_2R},\nonumber\\
p_2&=&\frac{a_1-a_3+2(a_2-a_3)\mu_1R}{\mu_1+\mu_2+2\mu_1\mu_2R}
\end{eqnarray}
and the electric mass formula (\ref{emeac}) gives
\begin{equation}
{\bf a}\cdot{\bf q}=g(a_2-a_1)p_1+g(a_3-a_2)p_2.
\end{equation}
Choosing $a_2=-a_1-a_3$ this corresponds to
\begin{equation}
{\bf a}=-2a_1\alphabf+2a_2\betabf.
\end{equation}
This agrees with the previous calculation \cite{LY}. To compare the
formulae directly, simply requires the substitution
\begin{eqnarray}
a_1&=&\xi s_1+\eta\mu_1,\nonumber\\
a_3&=&\xi s_3+\eta\mu_2.
\end{eqnarray}

\section{The $(2,[1])$ dyons in $SU(3)$ gauge theory \label{dsect}}
\news

\ns Let us apply the above formalism to the $(2,[1])$ case first
studied by Dancer \cite{D1,D2}. The metric is known in this case and
so the Tong formula could be used to calculate the electric mass,
however, it is easier to use the Nahm data formula we have derived.  A
$(2,[1])$-monopole is a 1/2-BPS configuration with two massive
monopoles and one massless one. The asymptotic form of the $b$ field
is
\begin{equation}
b=\frac{2\mu}{3}\frac{i}{2}\left(\begin{array}{ccc}-2&&\\&1&\\&&1\end{array}\right)-\frac{g}{12\pi r}\frac{i}{2}\left(\begin{array}{ccc}-2&&\\&1&\\&&1\end{array}\right).
\end{equation}
Thus, with $\alphabf$ and $\betabf$ the standard $SU(3)$ simple
roots, the magnetic charge is given by 
\begin{equation}
{\bf g}=g(2\alphabf+\betabf)
\end{equation}
and there are two massive $\alphabf$ monopoles and a massless
$\betabf$ monopole. The magnetic charge is purely abelian, because
${\bf g}\cdot \betabf=0$.

The Nahm data are over the interval $(-2\mu/3,\mu/3)$. In line with
common practice, this is translated to the interval $(0,\mu)$ and so
the data are 
\begin{equation}\label{2s1s}
\begin{array}{c}
\begin{picture}(205,65)(55,605)
\thinlines
\put( 60,620){\line( 1, 0){130}}
\put(100,620){\line( 0, 1){ 40}}
\put(100,660){\line( 1, 0){ 60}}
\put(160,660){\line( 0,-1){ 20}}
\put(160,640){\line( 1, 0){  2}}
\put(162,640){\line( 0,-1){ 20}}
\multiput(160,640)(0.00000,-8.00000){3}{\line( 0,-1){  4.000}}
\put( 95,623){\vector(0,1){  36}}  
\put( 95,623){\vector(0,-1){  0}} 
\put(162,620){\line( 0,-1){  5}}
\put(100,620){\line( 0,-1){  5}}
\put(160,620){\line( 0,-1){  5}}
\put( 84,638){\makebox(0,0)[lb]{\smash{\SetFigFont{12}{14.4}{rm}2}}}
\put( 97,605){\makebox(0,0)[lb]{\smash{\SetFigFont{12}{14.4}{rm}$0$}}}
\put(157,605){\makebox(0,0)[lb]{\smash{\SetFigFont{12}{14.4}{rm}$\mu$}}}
\put(167,623){\vector(0,1){  15}}
\put(167,623){\vector(0,-1){  0}}
\put(172,626){\makebox(0,0)[lb]{\smash{\SetFigFont{12}{14.4}{rm}1}}}
\end{picture}\end{array}
\end{equation}
There is no boundary condition at $s=\mu$.

The residual $SU(2)$ action on the monopole corresponds to gauge
inequivalent large gauge transformations of the Nahm data. The gauge
action is given by (\ref{Ndgact}) with $G$ an $SU(2)$ valued function
of $s$ with $G(0)={\bf 1}_2$. The gauge equivalence is given by the
small gauge transformations: those with $G(\mu)={\bf 1}_2$. Thus,
$G(\mu)$ parameterizes the $SU(2)$ action of large gauge
transformations. This action corresponds to the $SU(2)$ action on the
fields.

The Nahm equations are solved by
\begin{equation}
T_i(s)=-\frac{i}{2}\sigma_i f_i(s)
\label{nahmdata}
\end{equation}
where 
\begin{eqnarray}
f_1(s)&=& -\frac{D \cn{k}{Ds}}{\sn{k}{Ds}},\nonumber\\
f_2(s)&=&-\frac{D \dn{k}{Ds}}{\sn{k}{Ds}},\nonumber\\
f_3(s)&=&-\frac{D}{\sn{k}{Ds}}
\end{eqnarray}
are the well-known Euler top functions. There is a pole at $s=0$.
The data must be analytic for $s\in(0,\mu]$ and so $D<2K(k)/\mu$.
These solutions are acted on by the $SU(2)$ group action along with a
rotational $SO(3)$ action to give an eight-dimensional moduli space.

We now consider 1/4-BPS configurations.  Substituting these $(2,[1])$
Nahm data into the covariant Laplace equation gives the three Lam\'e
equations
\begin{equation}
\frac{d^2p_i}{ds^2}=(f_j^2+f_k^2)p_i
\end{equation}
where $(i\,j\,k)$ is a permutation of $(1\,2\,3)$ and
\begin{equation}
p=-\sum_i  \frac{i}{2}\sigma_i p_i.
\end{equation}
A two-parameter family of $p_i$ solving the relevant Lam\'e equation
is given by
\begin{equation}
p_i(s)=\alpha_i f_i(s)F_i(s)+\beta_i f_i(s)
\end{equation}
where 
\begin{equation}
F_i(s)=\int^s_{0}\frac{d s}{f_i(s)^2}.
\end{equation}

For $p$ to correspond to a 1/4-BPS state it must be finite and hence
$\beta_1=\beta_2=\beta_3=0$. This condition was also used to fix the
lower integration limit. $p$ is finite for any $\alpha$ since
$D<2K(k)/\mu$. Thus
\begin{equation}
{\bf a}\cdot{\bf
q}=\frac{g}{2}\sum_i\alpha_i^2\left(XF_i^2+F_i\right)
\label{aqdancer}
\end{equation}
where $f_i=f_i(\mu)$, $F_i=F_i(\mu)$ and $X=f_1f_2f_3$.

This expression needs to be normalized. $p(\mu)$ is in the algebra of the $SU(2)$ action on the moduli space and determines ${\bf a}$. If 
\begin{equation}
p(\mu)=\nu\sum_i n_i \frac{i}{2}\sigma_i
\end{equation}
with a unit vector ${\bf n}$, then 
\begin{equation}
{\bf a}\cdot{\bf q}=\frac{g}{2}\nu^2\sum_i n_i^2 c_i^2
\end{equation}
where
\begin{equation}\label{ciexp}
c_i=\frac{\sqrt{XF_i^2+F_i}}{f_iF_i}.
\end{equation}
These are the same as the $c_i$ appearing in the
metric\cite{D1,I}. $c_1$ is not normally written in this
form. However, it can be converted into it by using the integration by
parts identity
\begin{equation}
F_1+F_2+F_3+\frac{1}{X}=0.
\end{equation}

We can make spatial rotations and group rotations of the Nahm data.
Any $SU(2)$ gauge rotation changes the position of the massless monopoles.
After an $SU(2)$ gauge transformation of the  Nahm data (\ref{nahmdata}), 
the position of the massless position can be given by 
\begin{eqnarray}
{\bf r} &=&-i((T_1)_{22}, (T_2)_{22}, (T_3)_{22})  \nonumber \\
&=& (f_1 \sin{\theta} \cos{\varphi} , f_2\sin{\theta}\sin{\varphi}
, f_3 \cos{\theta}).
\label{ellipsoid}
\end{eqnarray}
Thus, the position of the massless monopole lies on an ellipsoid
\begin{equation}
\frac{x^2}{f_1^2} + \frac{y^2}{f_2^2}+ \frac{z^2}{f_3^2} = 1.
\end{equation}
The $p(t)$ for this gauge transformed Nahm data is simply the gauge
transformed version of the previous result. The $p(\mu)$ 
generates an infinitesimal $U(1)$ transformation  and so should leave the
position of the massless monopole invariant. Thus, 
\begin{equation}
{\bf n} = (\sin{\theta}\cos{\varphi}, \sin{\theta}\sin{\varphi},\cos{\theta})
\end{equation}
up to the sign. Therefore, we see that
the potential ${\bf a}\cdot {\bf q}$ depends on the position of the
massless monopole on the ellipsoid.

${\bf a}$ is determined by $p(\mu)$ and ${\bf a}\cdot{\bf q}$ is
known. Hence, the asymptotic value of the Higgs field $a$ is known. It
is
\begin{equation}
a\simeq
\frac{i}{2} \nu \; {\rm  block \, diag} \left(0,\sum_i n_i \sigma_i \right) 
\times \left( 1 - \frac{g\sum_i n_i^2 c_i^2}{8\pi r} \right).
\end{equation}
After diagonalizing, we get the Higgs expectation value
\begin{equation}
{\bf a} = \nu \betabf
\end{equation}
and the unbroken $U(1)$ charge is
\begin{equation}
{\bf q} = \frac{g\nu }{2} \sum_i n_i^2 c_i^2 \betabf.
\label{qdancer}
\end{equation}
Clearly, the electric charge of the 1/4-BPS configuration depends on
the position of the massless monopole.

\subsection{The field theory of the $(2,[1])$ example}

\ns In this subsection, we consider the behavior of the potential
calculated above and we describe the physics implied by this
behavior. It appears that the electric mass is minimized when the
massless $\betabf$ monopole is coincident with one of the $\alphabf$
monopoles and the two $\alphabf$ monopoles are infinitely
separated. In the next subsection, we calculate this minimum value
using string theory.

Quite a lot is known about the $(2,[1])$-monopole \cite{D1,D2,DL1,DL2,I}.
After fixing the center of mass and the overall phase, the moduli space
has an isometric $SU(2)$ action corresponding to the residual symmetry
and an $SO(3)$ action corresponding to rotation. These actions may be
fixed by assuming, as we did above, that $T_i$ is proportional to
$\sigma_i$ and that the top functions are ordered $f_1^2\le f_2^2\le
f_3^2$. This quotient space is two-dimensional and is parameterized by
$D$ and $k$. However, since the $SO(3)$ action is not free, this
quotient space is not a manifold.  A two-dimensional, totally geodesic
manifold which includes the quotient space was introduced in
\cite{DL1}. It is called $Y$ and is the manifold of Nahm data with
$T_i$ proportional to $\sigma_i$ but with no ordering assumption.

\begin{figure}[tb]
\begin{center}
\begin{picture}(290,250)(0,0)
\thinlines
\put(10,14){\epsfig{file=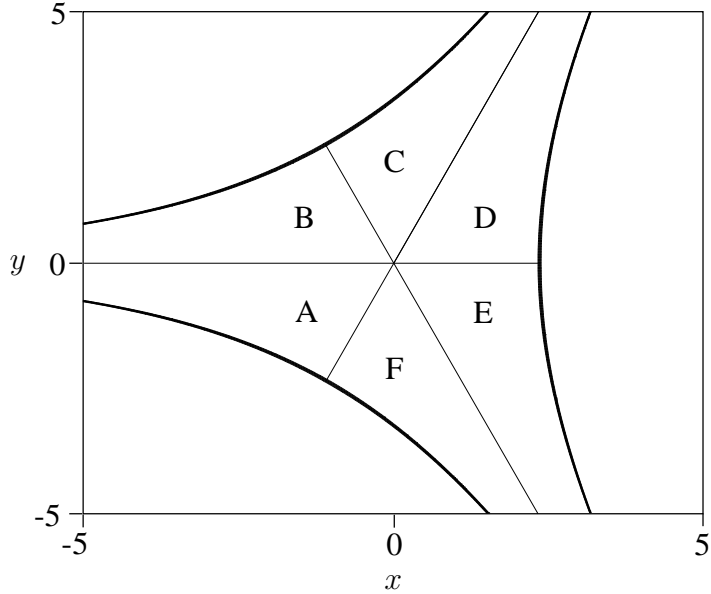,width=9cm}}
\put(0,134){$y$}
\put(157,0){$x$}
\end{picture}
\end{center}
\caption{The manifold $Y$.\label{Yfig}}
\end{figure}

$Y$ is pictured in Figure \ref{Yfig}. There are six identical regions
labeled A to F. Each is identical to the quotient space but with a
different ordering of the top functions. The coordinates on the space 
are 
\begin{eqnarray}
x&=&f_1^2+f_2^2-2f_3^2,\nonumber\\
y&=&\sqrt{3}(f_2^2-f_1^2),
\end{eqnarray}
so region B corresponds to the ordering $f_1^2\le f_2^2\le f_3^2$. The
thick boundary corresponds to $D\mu=2K$ and is a boundary of the
moduli space. On the boundary between regions, two of the top
functions are equal and the corresponding monopole is axially
symmetric. This can happen in two ways: either $k=0$ or $k=1$. When
$k=0$, the $(2,[1])$-monopole is torus shaped and coincident. This is
referred to as the trigonometric case, because the Euler top functions
are trigonometric. This is what happens, for example, on the boundary
between B and C.  When $k=1$, the $(2,[1])$-monopole may separate into
two individual monopoles. This is referred to as the hyperbolic case,
because the Euler top functions are hyperbolic; for example, on the
boundary between B and A:
\begin{eqnarray}
f_1(s)=f_2(s)&=&-D\mbox{cosech}\,Ds\nonumber\\
       f_3(s)&=&-D\coth{Ds}.
\end{eqnarray}
In this case, the two monopoles separate along the $x_3$-axis.  When
$D$ is large, it is the separation of the two monopoles
\cite{DL2,HIM}.  The clouds get bigger as the thick boundary is
approached and the $\alphabf$ monopoles separate down the legs
\cite{DL1,DL2,I}.

\begin{figure}[tb]
\begin{center}
\begin{picture}(290,250)(0,0)
\thinlines
\put(10,14){\epsfig{file=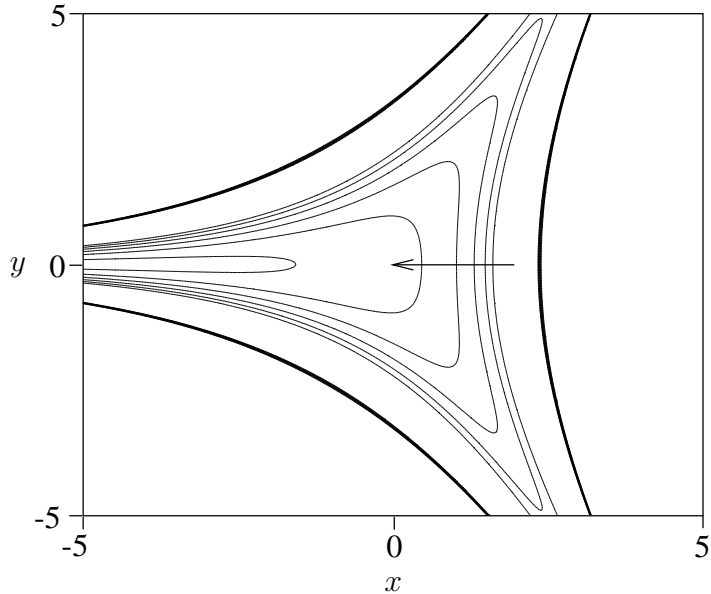,width=9cm}}
\put(0,134){$y$}
\put(157,0){$x$}
\end{picture}
\end{center}
\caption{A contour plot of $c$ as a potential on $Y$. The arrow points
  down the slope.\label{cYfig}}
\end{figure}

The expressions for the $c_i$'s are quite complicated. They are
plotted numerically in Figure \ref{cYfig}. In this figure, $c_3$ is
plotted in regions A and B, $c_2$ in C and F and $c_1$ in D and E.
This is done because the top functions are in a different order in
each region. Therefore, in each region, a different $c_i$ corresponds to
each $\sigma_i$. In Figure \ref{cYfig}, the $c_i$ plotted is the one
which corresponds to $\sigma_3$. The result is a continuous function
$c$ on $Y$.

$c$ seems to become infinitely large at the boundary. It seems to
increase steadily down the CD and EF legs and decrease down the AB
leg.  This can be confirmed by doing an explicit calculation with
$k=1$. In this case, it follows from the hyperbolic expressions that
\begin{equation}
c_3^2|_{k=1}=\frac{(\cosh{D\mu}\sinh{D\mu}-D\mu)D}
{D\mu\cosh{D\mu}\sinh{D\mu}-\mbox{sinh}^2\,D\mu}.
\end{equation}
This has a minimum for infinite $D$
\begin{equation}
\lim_{D\rightarrow\infty}{c_3^2|_{k=1}}=\frac{1}{\mu}.
\label{c3limit}
\end{equation}
Hence, the potential takes a minimum value for minimum cloud size and
maximum separation of the $\alphabf$ monopoles. This leads to a
critical electric charge which agrees with the string theory
(\ref{stq}). Similar calculations confirm that $c_1^2$ and $c_2^2$
diverge as $D$ goes to infinity.

This situation is similar to the $(1,[1],1)$ case discussed in
\cite{lee}. The uncharged monopole consists of two massive monopoles
and a massless one. As described in \cite{HIM}, the $SU(2)$ symmetry
acts on the position and charge of the massless monopole. This action
moves the massless monopole about on the ellipsoid known as the
massless cloud. If the $SU(2)$ symmetry is broken to $U(1)$, the
massless $\betabf$ seems to acquire a definite position.  The electric
1/4-BPS energy and the electric charge depend also on the position of
the massless monopole.

One of us $(KL)$, working with collaborators, has recently considered the
low energy effective lagrangian for 1/4-BPS dyons \cite{Betal2}. When
the electric part of the energy, which is always positive, is small,
then the 1/4-BPS configurations can be regarded as being very close to
1/2-BPS configurations. In fact, it may be regarded as being 
excited states of 1/2-BPS configurations. When the asymptotic value
${\bf a}$ is zero, only 1/2-BPS configurations appear and their low
energy dynamics are given by the moduli space metric.  When ${\bf a}$
is small compared to ${\bf b}$, the low energy effective lagrangian
is shown to be
\begin{equation}
L = \frac{1}{2} \sum_{p,q}g_{pq}(z) \dot{z}^p \dot{z}^q - U(z)
\end{equation}
where the potential is $U(z) = \frac{1}{2} {\bf a}\cdot {\bf q}$ and so 
\begin{equation}
U(z) = \sum_{p,q}\frac{1}{2} g_{pq}(z) ({\bf a}\cdot {\bf K}^p) ({\bf a}\cdot
{\bf K}^q).
\end{equation}

The potential appears because there are two Higgs fields involved in
the 1/4-BPS configuration and the electromagnetic force does not
exactly cancel the Higgs force. This effective lagrangian is of the order
\begin{equation}
L \sim v^2, \epsilon^2
\end{equation}
where $v$ is the order of velocities $\dot{z}^a$ and $\epsilon$ is the
order of $|{\bf a}|/|{\bf b}|$. This lagrangian is valid when $v\ll 1$ and 
$\epsilon \ll 1$. It has $N-1$ conserved electric charges,
one for each unbroken $U(1)$ symmetry. There is a BPS bound on this
newtonian lagrangian and field theoretic 1/4-BPS configurations appear
as BPS configurations.

When $\nu \ll \mu$, the electric contribution to the magnetic energy
is small if the electric charge is small.  In the $(2,[1])$ case, the
kinetic part of the low energy effective lagrangian is the metric on
the space of $(2,[1])$ monopoles: the Dancer metric \cite{D1}.  The
potential is
\begin{equation}
U(D,k) = \frac{g^2\nu^2}{2} \sum_i c_i^2 n_i^2, 
\end{equation}
that is, one half of the electric 1/4-BPS energy.  Since we are considering
only the relative motion, there is only one
conserved $U(1)$. The position of the $\betabf$ monopole lies on the
ellipsoid (\ref{ellipsoid}). In the trigonometric case $k=1$, the
massless monopole goes to the infinity when $D$ approaches its maximal
value $\pi/\mu$. In this limit, 
\begin{equation}
f_i \approx \frac{\pi}{\pi-\mu D}
\end{equation}
and so
\begin{equation}
c_i^2\approx \frac{\pi}{\mu(\pi -\mu D)}.
\end{equation}
Thus, the potential is linearly
increasing with the distance from the massless monopole to the two
$\alphabf $ monopoles. Therefore, the $\betabf$ monopole is confined. In the
hyperbolic case, with two massive monopoles  well-separated,
\begin{equation}
f_1=f_2\approx 0
\end{equation}
and 
\begin{equation}
f_3 \approx -D.
\end{equation}
In this limit, 
\begin{equation}
c_1^2 =c_2^2\approx D
\end{equation}
and 
\begin{equation}
c_3^2=1/\mu.
\end{equation}
Therefore, if we try to put the $\betabf$ monopole at
the middle of the line connecting two massive $\alphabf$ monopoles,
then the energy increases linearly with the distance. This again implies
that the $\betabf$ monopole should be confined to one of the two massive
$\alphabf$ monopoles.

\subsection{The string theory of the $(2,[1])$ example}

\ns In this subsection, we propose a string configuration corresponding
to the $(2,[1])$-monopole 1/4-BPS state and we justify this
proposal with the same sort of marginal stability argument as applied
to the $(1,1)$ dyon in \cite{LY}. 

It is believed that 1/4-BPS states correspond to configurations of
three-pronged strings \cite{B}. The string configuration given in
Figure \ref{11spic} is a well-understood example \cite{B,LY}. This
configuration of strings corresponds, in the field theoretic context, to
a $(1,1)$-monopole with electric charge $q\alphabf$. The string
configuration becomes unstable when the $(q,0)$-string has zero
length, that is, when the string junction coincides with the D3-brane
labeled (2) in Figure \ref{11spic}.

The tension of a $(q,g)$-string is $\sqrt{q^2+g^2}$ and by balancing
the forces at the string junction, it is simple to calculate the angles
between the strings \cite{B,LY}. The critical angle at which the
$(q,0)$-string has zero length is easily calculated from the D3-brane
positions. This means that the critical electric charge is known. In
\cite{LY}, the covariant Laplace equation (\ref{CL}) is solved for
$(1,1)$-monopoles with sufficient explicitness to allow the electric
charge to be calculated. It is found that the electric charge has a
maximum when the two monopoles are infinitely separated and that this
charge is equal to the critical charge calculated from the string
theory. Therefore, the critical charge can be calculated from either the
string theory or the field theory.

\begin{figure}[tb]
\begin{center}
\begin{picture}(120,115)(-70,-65 )
\thinlines
%
%
\put(0,0){\line( 1, 1){20}}
\put(0,0){\line( 1,-1){50}}
\put(0,0){\line(-5, 2){50}}
\put( -69,24){\makebox(0,0)[lb]{\smash{\SetFigFont{12}{14.4}{rm}$(1)$}}}
\put(  24, 24){\makebox(0,0)[lb]{\smash{\SetFigFont{12}{14.4}{rm}$(2)$}}}
\put(55, -59){\makebox(0,0)[lb]{\smash{\SetFigFont{12}{14.4}{rm}$(3)$}}}
\put(12,0){\makebox(0,0)[lb]{\smash{\SetFigFont{12}{14.4}{rm}$(q,0)$}}}
\put(29,-24){\makebox(0,0)[lb]{\smash{\SetFigFont{12}{14.4}{rm}$(0,g)$}}}
\put(-57,0){\makebox(0,0)[lb]{\smash{\SetFigFont{12}{14.4}{rm}$(q,g)$}}}
\put( -50, 20){\circle*{5}}
\put(  20, 20){\circle*{5}}
\put(  50,-50){\circle*{5}}
\end{picture}
\end{center}
\caption{This is the configuration discussed in \cite{LY}. The dots
are D3-branes and the lines are strings.\label{11spic}}
\end{figure}
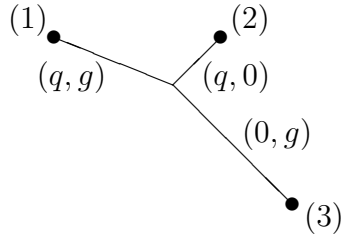

\begin{figure}[tb]
\begin{center}
\begin{picture}(265,165)(-15,-35 )
\thinlines
%
%
\put( 50,10 ){\line(0,1){60}}
\put( 50,70 ){\line(1,1){40}}
\put( 50,70 ){\line(-1,1){40}}
\put( 10,110){\circle*{5}}
\put( 90,110){\circle*{5}}
\put( 50,10 ){\circle*{5}}
\put( -9,117){\makebox(0,0)[lb]{\smash{\SetFigFont{12}{14.4}{rm}$(1)$}}}
\put( 30,0){\makebox(0,0)[lb]{\smash{\SetFigFont{12}{14.4}{rm}$(2)$}}}
\put(94,117){\makebox(0,0)[lb]{\smash{\SetFigFont{12}{14.4}{rm}$(3)$}}}
\put( 55,38){\makebox(0,0)[lb]{\smash{\SetFigFont{12}{14.4}{rm}$(0,2g)$}}}
\put( 78,82){\makebox(0,0)[lb]{\smash{\SetFigFont{12}{14.4}{rm}$(q,g)$}}}
\put( -9,82){\makebox(0,0)[lb]{\smash{\SetFigFont{12}{14.4}{rm}$(q,g)$}}}
\put( 42,-30){\makebox(0,0)[lb]{\smash{\SetFigFont{12}{14.4}{rm}(a)}}}
%
%
\put( 210,10 ){\line(0,1){60}}
\put( 210,70 ){\line(1,1){40}}
\put( 210,70 ){\line(-1,1){40}}
\put( 170,110){\circle*{5}}
\put( 250,110){\circle*{5}}
\put( 210,10 ){\circle*{5}}
\put( 174,110){\vector( 1,0){72}}
\put( 174,110){\vector(-1,0){ 0}}
\put( 170,106){\vector(0,-1){96}}
\put( 170,106){\vector(0, 1){ 0}}
\put(206,113){\makebox(0,0)[lb]{\smash{\SetFigFont{12}{14.4}{rm}$\nu$}}}
\put(159, 60){\makebox(0,0)[lb]{\smash{\SetFigFont{12}{14.4}{rm}$\mu$}}}
\multiput( 210,70)(8,0){4}{\line(1,0){4}}
\qbezier[12]( 230,70)(230,79  )( 224,84)
\put(234,78){\makebox(0,0)[lb]{\smash{\SetFigFont{12}{14.4}{rm}$\theta$}}}
\put( 202,-30){\makebox(0,0)[lb]{\smash{\SetFigFont{12}{14.4}{rm}(b)}}}
\end{picture}
\end{center}
\caption{These two pictures are of the Y-shaped string configuration, dots
  denote D3-branes and lines denote strings. (a) shows which string
  is which and (b) shows the lengths $\mu$ and $\nu$ and the angle $\theta$.\label{21spic}}
\end{figure}
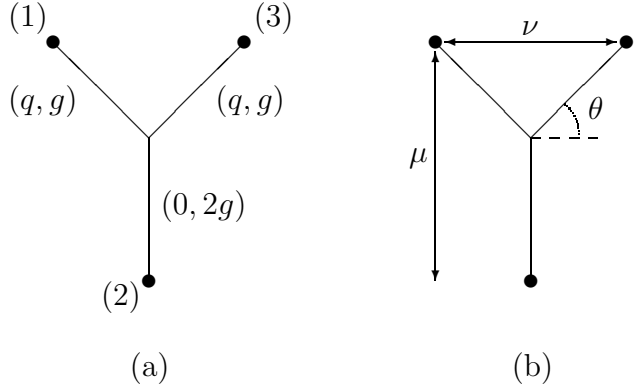

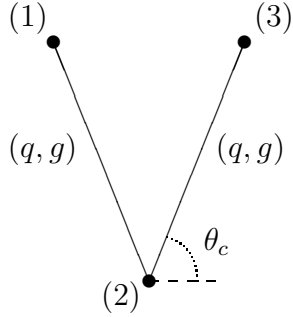
\begin{figure}[tb]
\begin{center}
\begin{picture}(115,130)(-15,0 )
\thinlines
%
%
\put( 50,10 ){\line(2,5){40}}
\put( 50,10 ){\line(-2,5){40}}
\put( -9,117){\makebox(0,0)[lb]{\smash{\SetFigFont{12}{14.4}{rm}$(1)$}}}
\put( 30,0){\makebox(0,0)[lb]{\smash{\SetFigFont{12}{14.4}{rm}$(2)$}}}
\put(94,117){\makebox(0,0)[lb]{\smash{\SetFigFont{12}{14.4}{rm}$(3)$}}}
\put( 78,62){\makebox(0,0)[lb]{\smash{\SetFigFont{12}{14.4}{rm}$(q,g)$}}}
\put( -9,62){\makebox(0,0)[lb]{\smash{\SetFigFont{12}{14.4}{rm}$(q,g)$}}}
\multiput( 50,10)(8,0){4}{\line(1,0){4}}
\qbezier[12]( 70,10)( 70,24)(57.4,28.5  )
\put(73,23){\makebox(0,0)[lb]{\smash{\SetFigFont{12}{14.4}{rm}$\theta_c$}}}
\put( 10,110){\circle*{5}}
\put( 90,110){\circle*{5}}
\put( 50,10 ){\circle*{5}}
\end{picture}
\end{center}
\caption{The V-shaped string configuration, this configuration is unstable.\label{21unstpic}}
\end{figure}

A similar calculation is undertaken in this section. A string configuration
is proposed and the electric mass is calculated when the configuration
is marginally stable.  The proposed string configuration is Y-shaped
and is shown in Figure \ref{21spic}.  The angle $\theta$ can be
calculated by balancing forces at the junction and is given by
\begin{equation}
\sin{\theta}=\frac{g}{\sqrt{q^2+g^2}}.
\end{equation}
This configuration becomes unstable if the $(0,2g)$-string has zero
length and the Y-shape degenerates to the V-shaped configuration
illustrated in Figure \ref{21unstpic}. At the onset of instability
\begin{equation}
\theta=\theta_c
\end{equation}
where
\begin{equation}
\sin{\theta_c}=
\frac{\mu}{\sqrt{\left(\frac{1}{2}\nu\right)^2+\mu^2}}.
\end{equation}
This means that the critical electric charge is $q_c$ where
\begin{equation}\label{stq}
q_c=\frac{g\nu}{2\mu}.
\end{equation}
{}From the string picture it would appear that this critical value of the
electric charge is a minimum. This contrasts with the $(1,1)$ case,
where the critical value is a maximum. The above value of the
critical charge is identical to the charge obtained from
(\ref{qdancer}) and (\ref{c3limit}). This critical value
corresponds to the 1/4-BPS electric charge of infinitely separated two
massive monopoles, with the massless monopole on top of either massive
monopole.

\begin{figure}[t]
\begin{center}
\begin{picture}(115,130)(-15,0 )
\thinlines
%
%
\put( 50,10 ){\line(2,5){40}}
\put( 50,10 ){\line(-2,5){40}}
\put( 10,110 ){\line(1,0){80}}
\put( 10,110){\circle*{5}}
\put( 90,110){\circle*{5}}
\put( 50,10 ){\circle*{5}}
\put( -9,117){\makebox(0,0)[lb]{\smash{\SetFigFont{12}{14.4}{rm}$(1)$}}}
\put( 30,0){\makebox(0,0)[lb]{\smash{\SetFigFont{12}{14.4}{rm}$(2)$}}}
\put(94,117){\makebox(0,0)[lb]{\smash{\SetFigFont{12}{14.4}{rm}$(3)$}}}
\put( 78,62){\makebox(0,0)[lb]{\smash{\SetFigFont{12}{14.4}{rm}$(0,g)$}}}
\put(
-8,62){\makebox(0,0)[lb]{\smash{\SetFigFont{12}{14.4}{rm}$(0,g)$}}}
\put(
32,119){\makebox(0,0)[lb]{\smash{\SetFigFont{12}{14.4}{rm}$(q,0)$}}}
\end{picture}
\end{center}
\caption{The $\nabla$-shaped string configuration, this
  configuration appears to have higher energy than the V-shaped configuration.\label{21false}}
\end{figure}
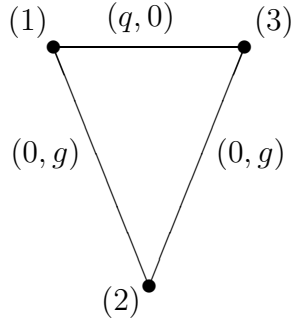

Figure \ref{21unstpic} might
create the suspicion that at some point the $Y$-shaped configuration
has larger energy than the $\nabla$-shaped configuration illustrated
in Figure \ref{21false}. Calculating the energies allays this
concern. The V-shaped configuration has energy
\begin{equation}
E_{\mbox{\scriptsize{V}}}=2g\mu\mbox{cosec}^2\,\theta=2g\mu(\mbox{cot}^2\,\theta+1)
\end{equation}
whereas, the $\nabla$-shaped configuration has energy
\begin{equation}
E_{\nabla}=2g\mu(\mbox{cot}^2\,\theta+\mbox{cosec}\,\theta).
\end{equation}
Therefore, $E_{\mbox{\scriptsize{V}}}\le E_\nabla$ with equality only
if $\theta$ is zero or $\pi/2$. Of course, calculations of this sort
are an over-simplification but they provide evidence favoring the
Y-shaped configuration over the $\nabla$-shaped one.

\section{Discussion\label{dissec}}
\news

\ns In this paper, we have derived a formula for the electric mass and
applied it to two examples. There are other examples that might also be
considered. It would be simple to extend the analysis to the
$(2,1)$-monopole, in which the $\betabf$ monopole has a magnetic
mass. In this case, the gauge orthogonality conditions require
\cite{HIM}
\begin{equation}
(\delta T_0)_{2,2}(s_2-)=\delta T_0 (s_2+).
\end{equation}
Other examples which might be considered are the
$([1],2,[1])$-monopole \cite{H,LL,HIM} or even the special higher
charge solutions discussed in \cite{HS5}.  It would also be
interesting to use the numerical ADHMN construction of \cite{HS1} to find
the $a$ field. This would reveal the spatial distribution of the
electric mass. This might be interesting in examples, like the one
considered in this paper, where there are monopoles with no magnetic
mass. It would also be instructive in examples, such as those in
\cite{HS5}, where there are extra minima of the Higgs field.

The dynamics of 1/4-BPS states are not fully understood. In the better
understood $(1,1)$ example the level set of the potential lies on a
group orbit. In the $(2,[1])$ example, this is not the case as  there are
monopoles with the same electric mass which cannot be group
transformed into each other. The geodesic motion on the $Y$ space was
studied by Dancer and Leese \cite{DL1}. It would be interesting to
determine how this motion is modified by the presence of the
potential.

\section*{Acknowledgment}
CJH warmly thanks the Physics Department and Center for Theoretical
Physics, Seoul National University for hospitality while part of this
work was undertaken. CJH thanks Fitzwilliam College, Cambridge for a
research fellowship and thanks Patrick Irwin and Paul Sutcliffe for
useful discussion. KL appreciates the Aspen Center for Physics,
Columbia University and PIMS of University of British Columbia for
support and hospitality.  KL is supported in part by the SRC program
of SNU-CTP, the Basic Science and Research Program under BRSI-98-2418,
and KOSEF 1998 Interdisciplinary Research Grant 98-07-02-07-01-5.

\end{document}